# О ВОЗМОЖНОСТИ СОЗДАНИЯ ТРОИЧНЫХ ЯЧЕЕК ПАМЯТИ НА ОСНОВЕ ПЕРФОРИРОВАННЫХ МАГНИТНЫХ ПЛЕНОК

## Е.Б. Магадеев, Р.М. Вахитов, Р.Р. Канбеков

В работе исследуются ферромагнитные пленки с сильной одноосной анизотропией типа «легкая плоскость» и обосновывается, что парные наноразмерные перфорации в таких пленках могут быть использованы в качестве ячеек памяти для записи и хранения данных в троичной системе исчисления. Изучена проблема считывания состояния ячеек такого типа, а также предложен подход к ее решению, заключающийся в измерении отклика системы на пикосекундный импульс внешнего магнитного поля. Получены параметры системы, при которых величина данного отклика оказывается наибольшей, а также проведены оценки этой величины как аналитическими, так и численными методами.

**Введение**

Одной из актуальных задач современной наноэлектроники является разработка методов сверхплотной записи информации на носителях, что приводит к необходимости развития новых физических принципов надежной энергонезависимой памяти. В этом отношении представляются перспективными попытки применения вихреподобных магнитных неоднородностей, разнообразие типов которых на сегодняшний день уже довольно велико (скирмионы, бимероны, цилиндрические магнитные домены и т.д. [1, 2]). Тем не менее, некоторые проблемы, связанные с устойчивостью и управляемостью структур такого типа, остаются до сих пор нерешенными, что затрудняет переход к их использованию на практике. В частности, концепция скирмионных кристаллов, когда магнитные неоднородности естественным образом образуют матрицу с наноразмерным пространственным периодом, является довольно привлекательной, однако даже теоретические расчеты показывают, что для возникновения таких текстур должны быть выполнены довольно специфические условия [3]. При этом экспериментальное наблюдение скирмионных кристаллов сталкивается с дополнительными сложностями. В качестве одного из возможных путей преодоления этих сложностей может быть предложено искусственное формирование пространственной матрицы за счет включения в магнитную пленку структурных дефектов или наноразмерных перфораций [4], также называемых антидотами [5]. Показано, что их наличие может привести к существенному росту плотности возникающих в пленке вихреподобных неоднородностей [6]. В то же время как механизм зарождения неоднородностей в этом случае, так и роль дефектов в соответствующих процессах зачастую остаются за рамками проводимых исследований. Напротив, в работах [7-10] был предложен и подробно изучен новый тип вихреподобных неоднородностей, которые возникают только в присутствии двух (или более) близкорасположенных антидотов при условии, что ферромагнитная пленка обладает сильной легкоплоскостной анизотропией, препятствующей выходу вектора намагниченности из плоскости пленки. Расчеты, приведенные в [10], показывают, что данное условие выполняется в довольно широком классе известных магнитных материалов; например, в $NdCo_5$ предложенные структуры остаются устойчивыми вплоть до диаметра перфораций 1 нм, чего достаточно для любых практических нужд. Также в [7] выявлено, что изучаемые неоднородности могут находиться в одном из трех (по меньшей мере) состояний, переключение между которыми осуществляется под воздействием электрического тока, пропускаемого через один из антидотов (динамические особенности этого процесса исследованы в [9]). Таким образом, вихреподобная неоднородность, локализованная в области пары перфораций, позволяет кодировать трит информации (в противовес биту), что открывает перспективы для значительного повышения плотности записи данных на носителе. Реализуемость такого подхода, очевидно, определяется наличием возможности не только переключать состояние описанной ячейки памяти, но также считывать это состояние. Разработке эффективного подхода к решению задачи считывания трита и посвящено данное теоретическое исследование.

**1. Физическая модель ячейки памяти**

Рассмотрим тонкую ферромагнитную пленку толщиной $h$, содержащую два одинаковых круглых антидота радиусом $R$ с центрами $O_1$ и $O_2$, как показано на рис. 1. При этом расстояние между центрами антидотов $|O_1O_2| = a$. Пусть материал образца характеризуется сильной одноосной анизотропией типа «легкая плоскость», так что вектор намагниченности $\boldsymbol{M} = M_s\boldsymbol{m}$ ($M_s$ – намагниченность насыщения, $\boldsymbol{m}$ – единичный вектор в направлении $\boldsymbol{M}$) практически не покидает плоскость пленки [10]. Тогда полную энергию магнетика можно записать в следующем виде [11]:

$$E = \int \left[ A\,(\nabla\theta)^2 - \boldsymbol{HM} - \tfrac{1}{2}\boldsymbol{H}_m\boldsymbol{M} \right] h\,dS, \qquad (1)$$

где $\theta$ – угол, определяющий ориентацию вектора $\boldsymbol{m}$ на плоскости, $A$ – обменный параметр, $\boldsymbol{H}$ – напряженность внешнего магнитного поля, $\boldsymbol{H}_m$ – напряженность размагничивающего поля, а интегрирование ведется по всей площади образца, не считая областей, занимаемых антидотами.

Пусть для начала внешнее поле отсутствует ($\boldsymbol{H} = 0$), а влияние размагничивающих полей пренебрежимо мало (в [8] показано, что данное условие выражается соотношением $2\pi M_s^2 \ll A/a^2$). Тогда в формуле (1) остается только член, отвечающий обменной энергии, и уравнение Эйлера-Лагранжа, представляющее собой условие минимума для соответствующего функционала, сводится к уравнению Лапласа $\Delta\theta = 0$. При этом граничные условия заключаются в требовании отсутствия нормальной компоненты grad $\theta$ на границах антидотов. Решения такого уравнения имеют вид [8]

$$\theta = k(\varphi_1 - \varphi_2), \qquad (2)$$

где $k$ – произвольное целое число, а $(r_1, \varphi_1)$, $(r_2, \varphi_2)$ – полярные системы координат, связанные с точками $C_1$ и $C_2$ (см. рис. 1), расстояние между которыми

$$|C_1C_2| = b = \sqrt{a^2 - 4R^2}. \qquad (3)$$

При $k = 0$ формула (2) описывает однородное распределение намагниченности. При $k = 1$ распределение имеет вид, показанный на рис. 2 (синие стрелки иллюстрируют реальное направление намагниченности, а серые – формальное продолжение решения (2) внутрь антидотов), причем неоднородность, отвечающая $k = -1$, очевидно, получается из него симметричным отражением. В [7, 9] показано, что переключение между обозначенными состояниями с $k = 0, \pm 1$ может осуществляться за счет воздействия магнитного поля проводников (в частности, состояние, показанное на рис. 2, может быть создано путем пропускания тока через левый антидот, так как циркуляция вектора $\boldsymbol{m}$ по охватывающему его контуру должна оказаться отличной от нуля). Возбуждение же состояний с $|k| > 1$ является гораздо более сложной задачей; кроме того, они обладают большей энергией [8] и худшей стабильностью [10], чем состояния с $k = 0, \pm 1$. Поэтому в дальнейшем будем рассматривать лишь три возможных состояния системы, которая представляет собой тем самым ячейку памяти, хранящую один трит информации.

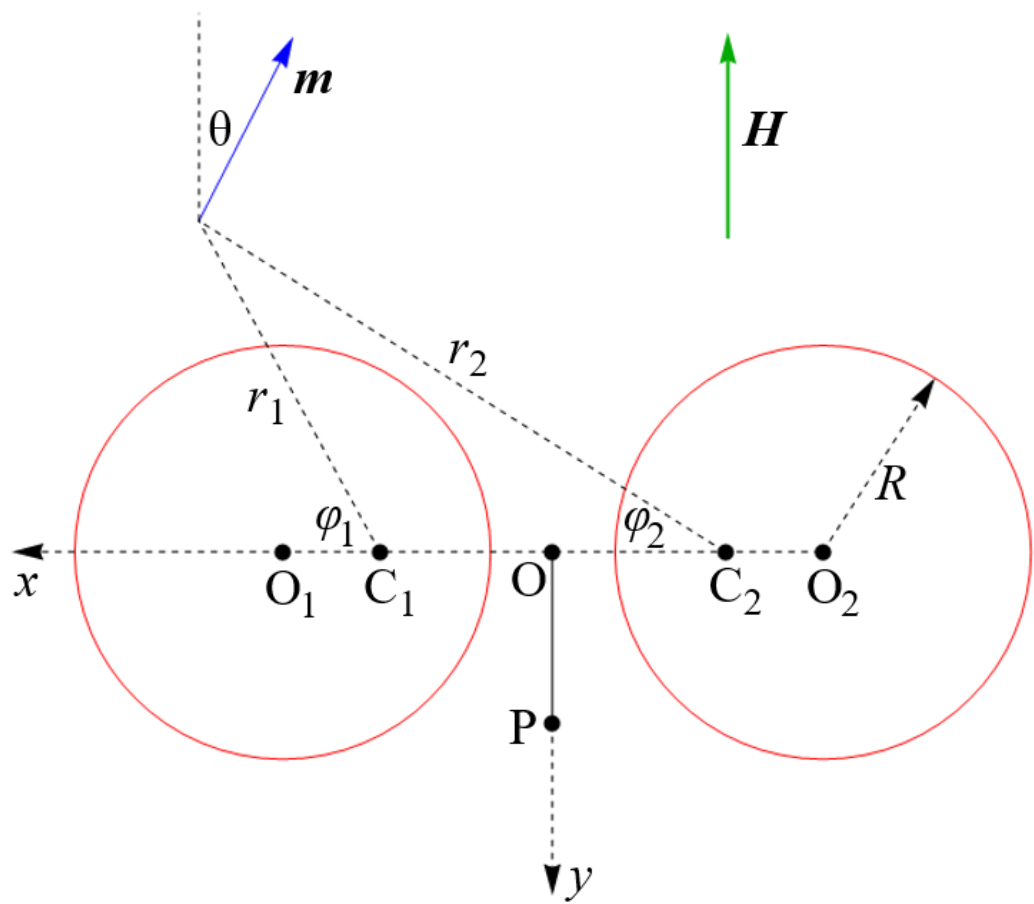

**Рис. 1. Геометрия системы**

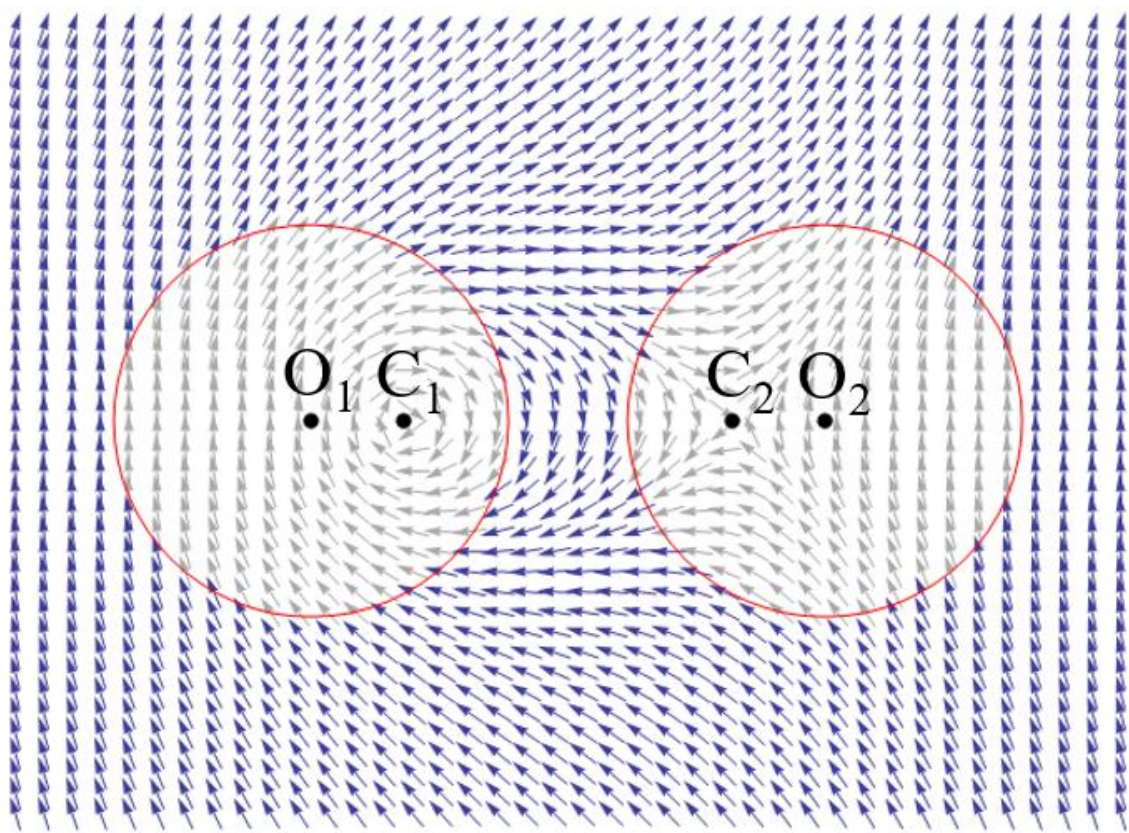

**Рис. 2. Распределение намагниченности при $k = 1$**

## 2. Подход к считыванию состояния ячейки

Введем декартову систему координат ($x$, $y$) так, как показано на рис. 1, и рассмотрим направления вектора намагниченности в различных точках оси O$y$ в состояниях с $k = 0, \pm 1$. Они схематически показаны на рис. 3 сплошными стрелками. Несложно видеть, что эти направления заметно различаются, вследствие чего непосредственное измерение намагниченности, например, в точке $y = 0.29b$ (на рис. 3 слева; в этой точке направления намагниченности для различных $k$ составляют между собой угол 120°) могло бы однозначно идентифицировать состояние системы. Тем не менее, данный подход не представляется практичным в условиях наличия большого числа ячеек памяти, включенных в единую матрицу. Заметим, однако, что в случае такой матрицы направление намагниченности на удалении от перфораций ($\theta = 0$) является единым для всех ячеек, а значит, кратковременное включение внешнего поля ***H*** в этом направлении (зеленая стрелка на рис. 1 и 3) приведет к тому, что все ячейки с равными значениями $k$ поведут себя одинаково, причем ячейки с $k = 0$ вообще не отреагируют на импульс. Тем самым становится возможно одновременно считать состояние всех ячеек, определенным образом отслеживая их отклик на подобное воздействие.

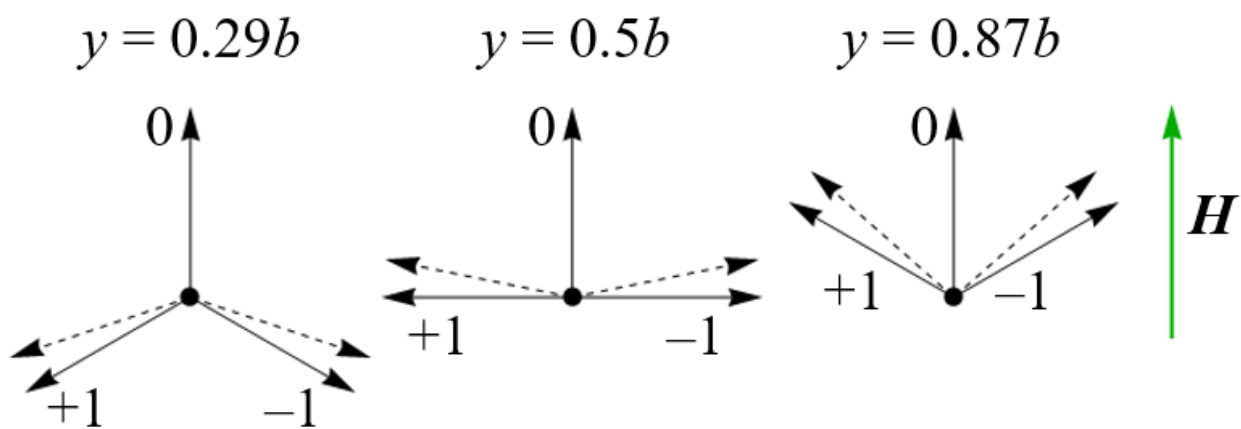

**Рис. 3. Зависимость направления намагниченности от величины *k* в различных точках на оси системы**

Рассмотрим проводящую прямоугольную рамку шириной $h$, расположенную перпендикулярно оси O$x$ в области $y_1 < y < y_2$, две стороны которой параллельны O$y$ и лежат на противоположных поверхностях пленки, а две другие проходят через пленку насквозь. ЭДС индукции $\varepsilon$ в такой рамке может быть найдена следующим образом:

$$\varepsilon = \varepsilon_0 + \varepsilon_m, \varepsilon_0 = \int_{y_1}^{y_2} 4\pi M_s \frac{\partial m_x}{\partial t}\Big|_{x=0} h\, dy, \varepsilon_m = \int_{y_1}^{y_2} \frac{\partial H_{mx}}{\partial t}\Big|_{x=0} h\, dy. \qquad (4)$$

Отсюда несложно видеть, что при включении поля ***H*** значения величины $\varepsilon$ для состояний с противоположными $k$ также будут противоположными. Кроме того, для состояния с $k = 0$ величина $\varepsilon$ будет равна нулю. Таким образом, $\varepsilon = k\, \varepsilon_1$ ($\varepsilon_1$ – фиксированная величина $\varepsilon$ при $k = 1$), то есть знак ЭДС, возникающей в рамке, фактически, можно отождествить со значением записанного трита, чем и обеспечивается его считывание.

Штриховые стрелки на рис. 3 показывают новые направления намагниченности через небольшое время после включения внешнего поля. Если $k = 1$, то значения компоненты $m_x$ в точках $y < b/2$ будут расти, а в точках $y > b/2$ – убывать. Поэтому в том случае, если $y_1 < b/2 < y_2$, области, расположенные по разные стороны от точки $y = b/2$ (точка P на рис. 1), будут, согласно (4), давать вклады разного знака в величину $\varepsilon_0$. Это нежелательно, поскольку наибольшая достоверность измерений может быть достигнута при наибольших абсолютных значениях измеряемой ЭДС. Следовательно, рамку целесообразно располагать либо между точками O и P (сплошная линия на рис. 1), при этом $\varepsilon_0 > 0$; либо «ниже» точки P, при этом $\varepsilon_0 < 0$.

Заметим, что слагаемым $\varepsilon_m$ в выражении (4), которое характеризует вклад размагничивающих полей, нельзя пренебрегать даже в том случае, когда эти поля не оказывают существенного влияния на структуру магнитной неоднородности. В связи с этим приведенное выше рассуждение, вообще говоря, не доказывает, что найденные положения рамки позволяют максимизировать также и абсолютную величину полной ЭДС $\varepsilon$. Тем не менее, из дальнейших расчетов станет ясно, что сделанные оценки оказываются вполне корректными благодаря существованию тесной связи между распределениями $\boldsymbol{M}$ и $\boldsymbol{H}_m$.

### 3. Аналитический расчет отклика ячейки

Чтобы рассчитать величину ЭДС, возникающей в рамке при включении поля $\boldsymbol{H}$, достаточно определить значения производных по времени, входящих в выражение (4), в момент времени $t = 0$. В [9] показано, что при наличии сильной легкоплоскостной анизотропии динамика магнитной структуры может быть описана уравнением движения следующего вида (оно совпадает по форме с одним из уравнений Ландау-Лифшица-Гильберта в угловых переменных при условии сильного затухания):

$$\frac{\partial \theta}{\partial t} = -\frac{\gamma}{\alpha M_s} \frac{\delta E}{\delta \theta}. \tag{5}$$

Здесь $\gamma$ – гиромагнитное отношение, а $\alpha$ – параметр диссипации. Поскольку перед включением поля система находится в состоянии равновесия, то в начальный момент времени ненулевой вклад в правую часть соотношения (5) дает только слагаемое в энергии (1), связанное с влиянием поля $\boldsymbol{H}$. Отсюда для $m_x = -\sin\theta$ и $m_y = -\cos\theta$ имеем:

$$\frac{\partial m_x}{\partial t} = \frac{\gamma H}{\alpha} \sin\theta \cos\theta, \frac{\partial m_y}{\partial t} = -\frac{\gamma H}{\alpha} \sin^2\theta. \tag{6}$$

При этом напряженность $\boldsymbol{H}_m$ размагничивающего поля, как и ее производная по времени, может быть найдена из уравнений Максвелла [11], так что

$$\text{rot}\,\frac{\partial \boldsymbol{H}_m}{\partial t} = 0, \text{div}\left(\frac{\partial \boldsymbol{H}_m}{\partial t} + 4\pi M_s \frac{\partial \boldsymbol{m}}{\partial t}\right) = 0. \tag{7}$$

Пусть рамка расположена существенно «ниже» точки P, так что $y_1 \gg b$. Тогда динамику системы достаточно рассмотреть на большом удалении от антидотов, где $r = r_1 \approx r_2 \gg b$, $\varphi = \varphi_1 \approx \varphi_2$. В этом случае из (2) (по-прежнему предполагая, что размагничивающие поля не оказывают заметного влияния на структуру неоднородности) для $k = 1$ приближенно имеем:

$$\theta = \frac{b \sin\varphi}{r}. \tag{8}$$

Подставляя (8) в формулы (6) и решая уравнения (7), получаем:

$$\frac{\partial m_x}{\partial t} = \frac{\gamma H b}{\alpha} \frac{\sin\varphi}{r}, \frac{\partial m_y}{\partial t} = 0, \frac{\partial H_{mx}}{\partial t} = \frac{2\pi M_s \gamma H b}{\alpha} \frac{\sin\varphi \cos 2\varphi}{r}, \frac{\partial H_{my}}{\partial t} = \frac{2\pi M_s \gamma H b}{\alpha} \frac{\cos\varphi \cos 2\varphi}{r}. \tag{9}$$

Используя соотношения (9) и учитывая, что $\varphi = 3\pi/2$ для точек, принадлежащих области рамки, из (4) окончательно имеем:

$$\varepsilon = \varepsilon_0/2 = -\varepsilon_m = -\frac{2\pi M_s \gamma H b h}{\alpha} \ln\frac{y_2}{y_1}. \tag{10}$$

Из выражения (10) следует, что абсолютная величина измеряемой ЭДС неограниченно растет с увеличением $y_2$, то есть теоретически мы могли бы сколь угодно усилить измеряемый отклик ячейки, выбирая рамки все большей длины. Разумеется, на практике данное обстоятельство не будет иметь места, поскольку при сколько-нибудь заметном удалении от антидотов отличие распределения намагниченности от однородного будет определяться уже не соотношением вида (8), а совокупностью случайных факторов.

Пусть далее рамка расположена точно между точками O и P, то есть $y_1 = 0$, $y_2 = b/2$. В этом случае получение явного решения уравнения (7) является затруднительным. Тем не менее, по аналогии с выражением (10) мы можем предположить, что соотношение $\varepsilon = \varepsilon_0/2$ по-прежнему приближенно выполняется. Тогда при $x = 0$ из (2) для $k = 1$ имеем

$$\theta = -2\arctan\frac{b}{2y} \tag{11}$$

и, подставляя соотношение (11) в первую формулу (6), из (4) получаем:

$$\varepsilon = \varepsilon_0/2 = \frac{\pi M_s \gamma H b h}{\alpha}(1 - \ln 2). \tag{12}$$

Сравнивая выражения (12) и (10), можно заметить, что абсолютные значения ЭДС для этих двух случаев оказываются равными, если в (10) $y_2/y_1 \approx 1.17$. Хотя данное отношение довольно невелико, это отнюдь не означает, что рамка длины $b/2$, размещенная «ниже» точки P, могла бы регистрировать гораздо более сильный отклик, чем рамка, помещенная между точками O и P, так как выражение (10) справедливо только при $y_1 \gg b$. В действительности расчет показывает, что отклик, противоположный ЭДС (12), регистрируется рамкой с параметрами $y_1 = 0.5b$, $y_2 \approx 0.99b$, которая имеет практически ту же длину, что и |OP|. Поэтому размещение рамки между точками O и P в итоге представляется оптимальным вследствие того, что в этом случае ячейка памяти оказывается наиболее компактной.

## 4. Численное моделирование отклика ячейки

Вывод выражения (12) опирался на целый ряд допущений; кроме того, полученные результаты относятся к значению $\varepsilon$ в момент времени $t = 0$ и потому не позволяют судить о зависимости ЭДС от времени. По этим причинам отклик ячейки (находящейся в состоянии с $k = 1$) на импульс внешнего поля был рассчитан также с помощью пакета микромагнитного моделирования OOMMF [12]. При этом рассматривался образец размерами 200 нм × 100 нм × 10 нм, содержащий два круглых антидота радиусом $R = 10$ нм и характеризующийся следующими значениями материальных параметров: $A = 10^{-12}$ Дж/м, $M_s = 10^4$ А/м, $\alpha = 0.1$, абсолютная величина константы одноосной анизотропии $10^7$ Дж/м$^3$. Продолжительность $\tau$ импульса внешнего поля напряженностью $H = 10^5$ А/м (эквивалентно 126 мТл) принималась равной 0.5, 1 или 2 пс. Также варьировалось расстояние между центрами антидотов: $a = 3R$ или $a = 4R$. В обоих случаях рамка располагалась таким образом, чтобы она охватывала все элементы расчетной сетки (в форме кубиков объемом 1 нм$^3$), дающие положительный вклад в значение начальной ЭДС $\varepsilon(t = 0)$. При этом были получены значения $y_1 = 1$ нм, $y_2 = 11$ нм для $a = 3R$ и $y_1 = 1$ нм, $y_2 = 19$ нм для $a = 4R$. Расчет длины отрезка |OP| = $b/2$ в соответствии с формулой (3) дает для этих случаев результаты 11.2 нм и 17.3 нм соответственно, то есть, действительно, $y_2 \approx b/2$. Это подтверждает ранее сделанный вывод о том, что положение рамки между точками O и P является оптимальным.

На рис. 4 показаны полученные зависимости $\varepsilon(t)$ при различных значениях параметров $a$ и $\tau$. Величина начальной ЭДС $\varepsilon(t = 0)$, которая не зависит от продолжительности импульса τ, оказалась равна 49.8 нВ для *a* = 3*R* и 109.4 нВ для *a* = 4*R*, притом что расчеты с использованием формулы (12) дают 47.7 нВ и 73.8 нВ соответственно. Таким образом, формула (12) позволяет получить довольно неплохие оценки ЭДС, точность которых, по всей видимости, снижается при больших расстояниях *a* между центрами антидотов.

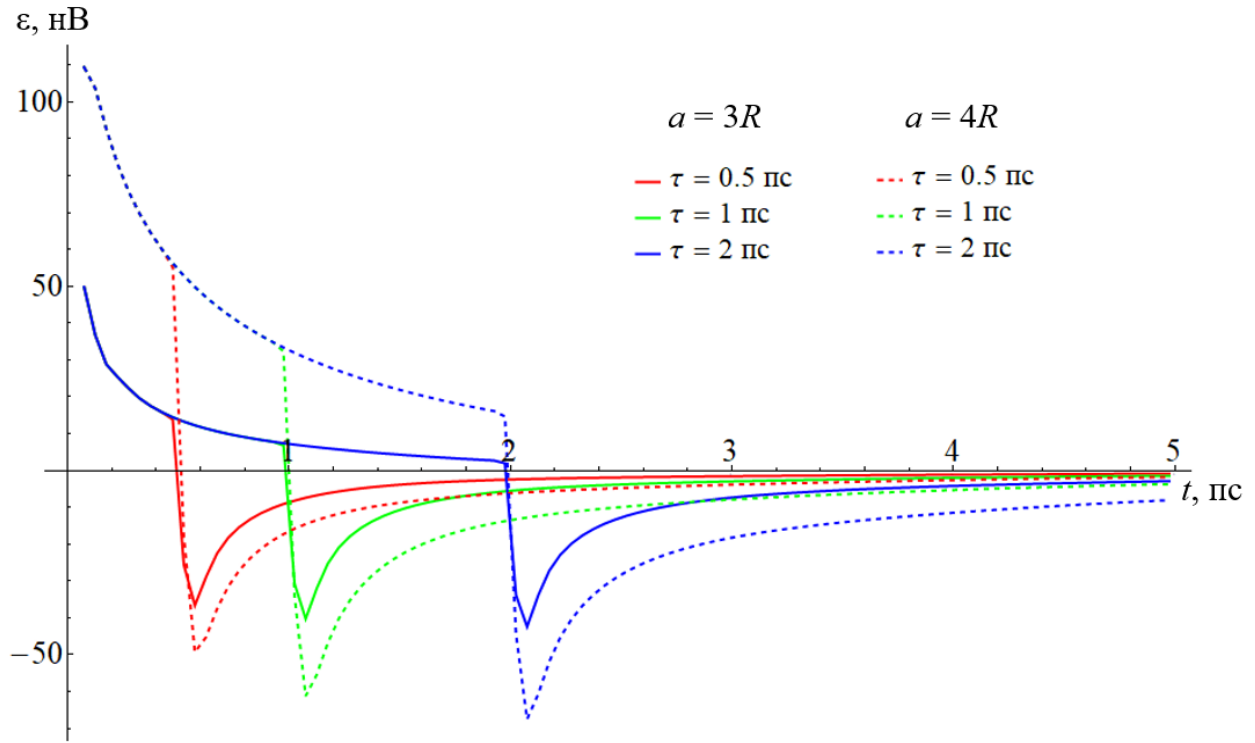

**Рис. 4. Результат численного моделирования зависимости ЭДС от времени**

Анализируя зависимости на рис. 4, несложно заметить, что отклик изучаемой системы на качественном уровне совпадает с откликом динамической системы первого порядка с одной степенью свободы на кратковременное внешнее воздействие. Действительно, если рассмотреть уравнение движения вида

$$f + T\frac{df}{dt} = g(t), \qquad (13)$$

где *f* – обобщенная координата системы, *T* – постоянный параметр, а функция *g*(*t*) принимает постоянное ненулевое значение при *t* < τ и обращается в нуль при *t* > τ, то при условии *f*(0) = 0 график функции ε = d*f*/d*t* будет иметь вид, аналогичный кривым на рис. 4. Ясно тем самым, что величина *f* может быть отождествлена с магнитным потоком через рамку (точнее, с его отличием от значения потока в равновесном состоянии), функция *g* – с влиянием внешнего поля, которое заключается в появлении нового равновесного значения потока *f*, а второе слагаемое в левой части уравнения (13) – с диссипативным членом, обеспечивающим релаксацию системы в соответствии с уравнением (5). Важным следствием из этой упрощенной модели является то, что помимо продолжительности импульса τ динамика изучаемых ячеек обусловлена единственным характерным временем, а именно – временем релаксации *T*. Эта величина не может зависеть от напряженности поля *H*, а также, судя по результатам, представленным на рис. 4, слабо зависит от расстояния *a* между центрами антидотов. Поэтому можно предположить, что в системе единиц СИ

$$T \sim \frac{\mu_0 M_S \alpha R^2}{\gamma A}, \qquad (14)$$

где $\mu_0$ – магнитная постоянная. Для значений параметров, использованных при моделировании, выражение в правой части (14) равно 0.57 пс, что довольно неплохо согласуется с динамикой, наблюдаемой на рис. 4. Таким образом, коэффициент пропорциональности в формуле (14) имеет значение, близкое к 1. Заметим, что, согласно этой формуле, миниатюризация ячеек памяти (уменьшение радиуса антидотов $R$) будет сопровождаться снижением времени релаксации, то есть повышением их быстродействия.

**Заключение**

Таким образом, мы показали, что парные перфорации ферромагнитной пленки, снабженные рамкой, которая расположена в области перфораций и имеет длину, сопоставимую с расстоянием между ними, может быть использована в качестве ячейки памяти, обеспечивающей запись, хранение и считывание трита информации. При этом значение трита, определяемое тем, в каком из топологически различных состояний находится магнитная структура ячейки, может быть отождествлено со знаком ЭДС, возникающей в рамке при включении пикосекундного импульса внешнего магнитного поля. Важной особенностью такого подхода к считыванию состояния ячейки является возможность считать записанные триты одновременно из большого числа ячеек, включенных в единую матрицу. Параметры как создаваемого импульса поля, так и отклика ячеек на этот импульс находятся, насколько можно судить из проведенных расчетов, в соответствии с современными техническими возможностями наноэлектроники, а также с требованиями, предъявляемыми к надежности и быстродействию записывающих устройств.

Следует отметить, что условие наличия сильной одноосной анизотропии типа «легкая плоскость», которое является существенным в рамках изложенной концепции, в действительности не налагает серьезных ограничений на классы магнитных материалов, применимых для создания троичных ячеек. Предварительные оценки показывают, что эффективной легкоплостной анизотропии, возникающей в тонких пленках вследствие влияния размагничивающих полей, в широком ряде случаев может оказаться достаточно, чтобы стабилизировать описанные выше вихреподобные структуры. Очевидно, что это открывает перспективы для создания предлагаемых ячеек памяти из доступных и сравнительно недорогих материалов (например, пермаллоев). Тем не менее, данный вопрос требует более тщательного изучения, и окончательные результаты будут представлены в ближайшем будущем.